# Outbursts Upon Cooling of Low-Temperature Binary Mixtures: Experiments and Their Planetary Implications

S. M. Raposa[1,2][†], A. E. Engle[1,2][†], S. P. Tan[3], W. M. Grundy[2,1], J. Hanley[2,1], G. E. Lindberg[4], O. M. Umurhan[5,6,7,8], J. K. Steckloff[3,9], C. L. Thieberger[1,2], S.C. Tegler[1]

[1] Department of Astronomy and Planetary Science, Northern Arizona University, Flagstaff, AZ, USA

[2] Lowell Observatory, Flagstaff, AZ, USA

[3] Planetary Science Institute, Tucson, AZ, USA

[4] Department of Chemistry and Biochemistry, Northern Arizona University, Flagstaff, AZ, USA

[5] Carl Sagan Center, SETI Institute, Mountain View, CA, USA

[6] Space Sciences Division, Planetary Systems Branch, NASA Ames Research Center, Moffett Field, CA, USA

[7] Department of Astronomy, Cornell University, Cornell, NY, USA

[8] Earth and Planetary Sciences, University of California, Berkeley, Berkeley, CA, USA

[9] Department of Aerospace Engineering and Engineering Mechanics, The University of Texas at Austin, Austin, TX, USA

† First authors

Corresponding authors: Shaelyn Raposa (sraposa@lowell.edu); Anna Engle (aengle@lowell.edu)

**KEY POINTS**

- Three binary systems exhibited outbursts of volatile materials due to freezing-induced overpressurization
- Solid-liquid-vapor curves can be used to predict when outbursts occur
- Outbursts have implications for explosive/effusive processes throughout the solar system and likely occur in many other volatile mixtures

**ABSTRACT**

For many binary mixtures, the three-phase solid-liquid-vapor equilibrium curve has intermediate pressures that are higher than the pressure at the two pure triple points. This curve shape results in a negative slope in the high-temperature region near the triple point of the less volatile component. When freezing mixtures in the negative slope regime, fluid trapped below confined ice has latent heat released with more vapor upon cooling, and thus increases in pressure. If the rising pressure of the confined fluid overcomes the strength of the confining solid, which may be its own ice, it can produce an abrupt outburst of material and an increase in the system's overall pressure. Here, we report experimental results of freezing-induced outbursts occurring in the $N_2/CH_4$, $CO/CH_4$, and $N_2/C_2H_6$ systems, and provide insight into the phenomenon through a thermodynamics perspective. We also propose other binary systems that may experience outbursts and explore the geological implications for icy worlds like Titan, Triton, Pluto and Eris, as well as rocky bodies, specifically Earth and Mars.

**PLAIN LANGUAGE SUMMARY**

The solar system is host to a number of active worlds, whose subsurface materials can well up and breach the surface through explosive and effusive means. Typically, these processes are initiated when the substance is heated and consequently pressurized, leading to the affected materials finding an upward means of escape. Interestingly, we have seen in the laboratory that mixtures can undergo overpressurization as a result of freezing. In this paper, we report our findings in experiments with mixtures of nitrogen/methane, carbon monoxide/methane, and nitrogen/ethane and refer to the process of overpressurized material breaching the surface during freezing as "outbursts." We represent these events through the use of phase diagrams, which are useful tools for understanding surface processes as they indicate where phase changes happen and what phases (e.g., vapor, liquid, solid, or a combination of these) will be most stable for any temperature, pressure, and composition conditions. It is likely that this outburst process occurs in many mixtures beyond those explored in the experiments in this paper. We discuss more mixtures that could have outbursts, and give examples of how this process is relevant on icy worlds like Titan, Triton, Pluto, and Eris; and rocky bodies, specifically Earth and Mars.

# 1. INTRODUCTION

The increasing pressure of gas associated with a liquid upon heating is familiar to everyone who has boiled a kettle for tea. It is fundamental to numerous industrial and geological processes, from steam engines and power-generating gas turbines to the geysers of Yellowstone National Park (Wyoming, U.S.A.). This paper introduces a somewhat less intuitive increase of gas pressure that can occur upon cooling and freezing of liquid mixtures and explores the potential geological implications of the outburst phenomenon. Of particular interest are situations where some volatile-containing fluid becomes trapped as it freezes such as within its own ice in a three-phase solid-liquid-vapor (SLV) equilibrium. The composition of the residual liquid evolves over time upon cooling, becoming more enriched in the volatile species as ice continues to form together with the increasing amount of vapor phase while releasing a net amount of latent heat. Therefore, as the cooling proceeds along the SLV equilibrium, the pressure imposed on the fluid increases and can trigger an outburst if it is high enough to overcome the strength of the confining solid. Once the frozen lid has been breached, gas bubbles are able to percolate to the surface, pushing on the slurry and forming mounds or even abrupt outbursts while escaping. The thermodynamics of the above situations is discussed in Appendix A.

In our previous work (Raposa et al., 2022) on some binary mixtures, part of the SLV equilibrium curve, where freezing occurs, has a negative slope, which means increasing pressure upon cooling. Most of the curves have maximum pressure at intermediate temperatures and compositions, which is higher than pressure at either of the pure substance triple points as shown in Figure 1. The topology of this SLV curve belongs to α-type in a recent classification for phase diagrams of binary mixtures (Quinzio et al., 2023). If the binaries exhibit two liquid phases at high pressures, then the pressure increases up to the quadruple point where 4 phases (SLLV, i.e., SLV plus another liquid) are in equilibrium, instead of having a continuous curve with a maximum, as shown in Figure 2. This latter topology belongs to β-type in the classification. In general, the topology is also determined by the degree of asymmetry of the mixture. The β-type SLV, for example, is found in more asymmetric binary mixtures than those with α-type SLV (Quinzio et al., 2023). The asymmetry may be associated with large differences in molecular weights or densities of the mixture components, low solid miscibility, low mutual solubility of liquids, etc.

We were specifically studying low-temperature liquids involving nitrogen ($N_2$), carbon monoxide (CO), methane ($CH_4$), and ethane ($C_2H_6$)—materials that occur in outer solar system settings, whereas the studied binary mixtures relevant to the current work are $N_2/CH_4$ and $CO/CH_4$, both of which show moderate solid miscibility and α-type SLV (Figure 1), as well as $N_2/C_2H_6$, which shows barely miscible solid solutions and the appearance of a second liquid, thus β-type SLV (Figure 2). From the study, we found that the outbursts happened only at the part of the SLV curves with negative slope near the less-volatile component, which is a common property of binary mixtures, e.g., the binary carbon dioxide/methane ($CH_4/CO_2$) system as also shown in Figure 1 for α-type SLV or methane/hydrogen sulfide ($CH_4/H_2S$) in Figure 2 for β-type SLV.

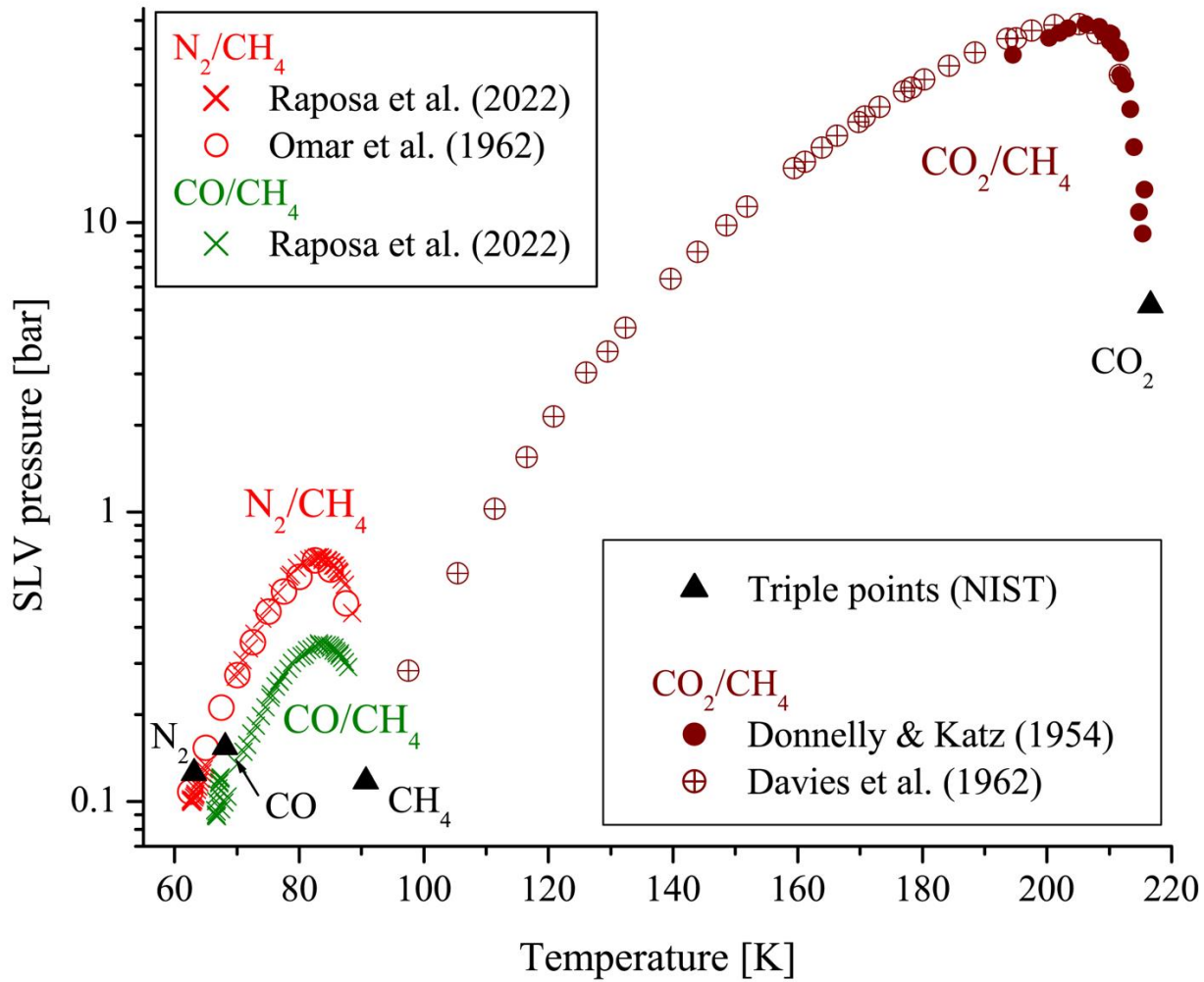

*Figure 1.* *The experimental freezing points at α-type solid-liquid-vapor (SLV) equilibrium for some binary mixtures that show maximum pressure.*

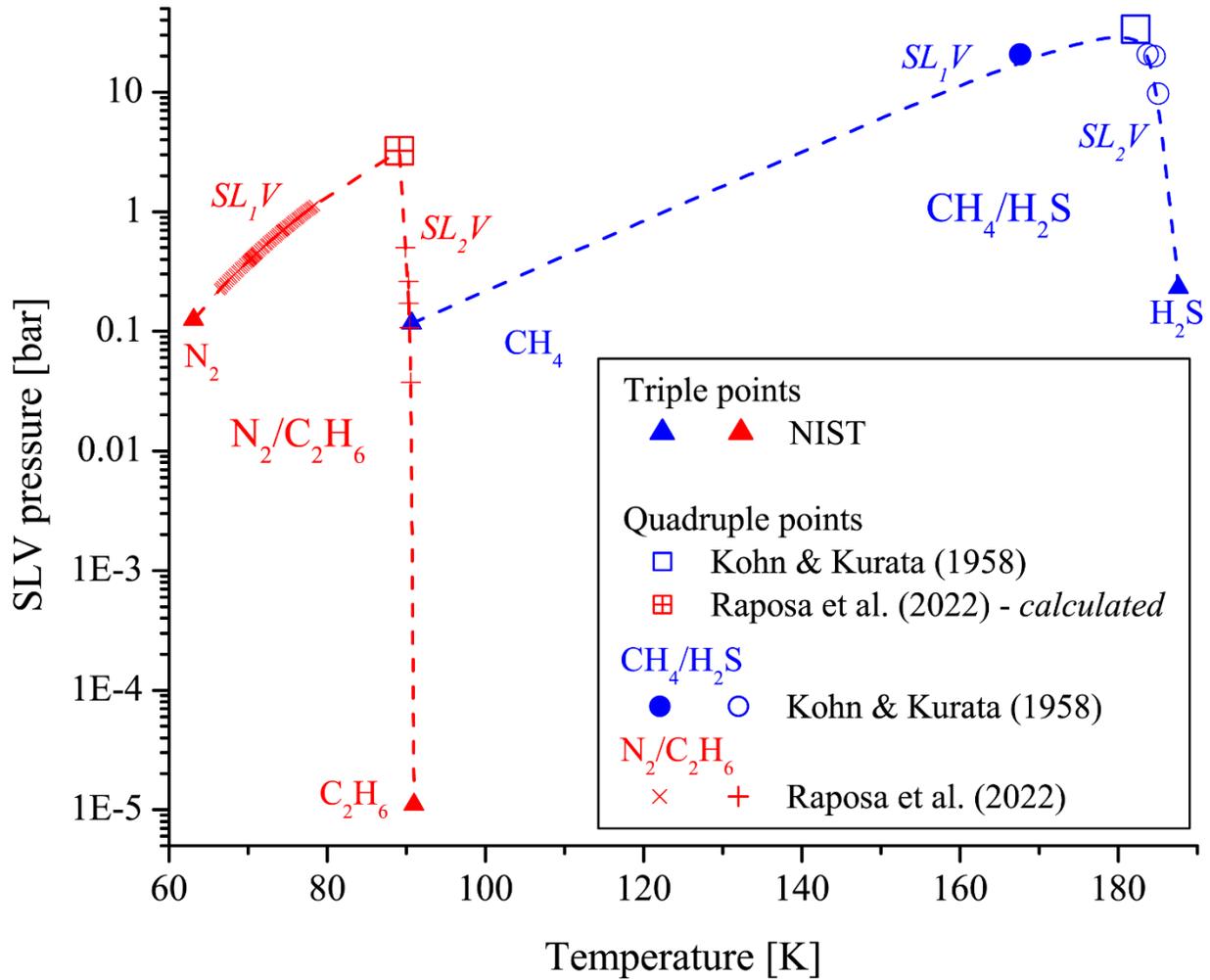

*Figure 2. The experimental freezing points at β-type SLV equilibrium for binary mixtures that have quadruple point SLLV. $L_1$ and $L_2$ are different liquid phases rich with the first and second species in the mixtures, respectively. The dashed curves are guides for eyes.*

## 2. METHODS

### 2.1. Experimental Set-up

The experiments described in this paper were conducted in the Astrophysical Materials Laboratory at Northern Arizona University. Experiments take place in a tiny, isolated environment system, where we can study liquid and ice mixtures down to 30 K. The materials to be studied are injected into the aluminum alloy sample cell, which has a cylindrical interior dimension of 15 mm in diameter and 20 mm in length. The cell is fitted with two sapphire windows that are forced against indium wire gaskets and allow us to view the sample by eye. The ability to view directly into the cell is a useful feature that provides us the ability to observe and monitor the outbursts.

Prior to the start of an experiment, the cell is cooled using a two-stage CTI 1050 closed-cycle helium refrigerator. Once the starting temperature is reached, the sample is mixed using materials from gas cylinders hooked up to stainless steel tubing. Gasses may be pre-mixed in a 0.5 L mixing chamber prior to injection into the cell or injected directly into the cell in small bursts. We aim to fill the cell halfway so there is a free upper surface to observe the outbursts. Preliminary experiments were carried out to determine how much of each species was needed to reach the desired fill capacity since the exact amount of material needed varies between binary systems. A pair of nichrome heaters are positioned above and below the cell. The temperature is controlled using the bottom heater, which creates a small vertical thermal gradient throughout the cell of <1 K. Since the conductivity of volatile ices is relatively low in relation to the cell materials, the parts of the sample closest to the cell walls and windows respond to temperatures changes first, with the interior of the sample responding more slowly, so we hold at each temperature step for at least 5 minutes until the sample stabilizes.

The reported measurement errors on our temperatures are ±0.5 K, mostly potential systematic uncertainties (random temperature excursions are less than 0.05 K). Temperature calibration tests were performed using pure $N_2$ and $CH_4$ in equilibrium at their triple points. Temperature offsets between DT-670 diode readings and the known phase change temperatures were corrected using a linear correction fitted to the measured triple point temperatures of $N_2$ and $CH_4$, plus the known temperature of the $N_2$ α-β ice phase transition, which occurs at 35.61 K (Scott, 1976). Pressure is monitored near the mixing volumes using a MKS Baratron model 627 F heated capacitance manometer, which measures a range of 5-5000 Torr. Pressures are reported with a precision of ±0.25% of the measured pressure.

For more information about the laboratory set-up, see previous publications (Ahrens et al., 2018; Grundy et al., 2011; Tegler et al., 2010, 2012, 2019).

**2.2. Experimental Protocol**

The sample is condensed into the cell once it has reached a temperature just above the freezing point of $CH_4$ or $C_2H_6$, ensuring the starting mixture is in liquid-vapor equilibrium. The less volatile species ($CH_4$ or $C_2H_6$) is injected into the cell first until it is roughly halfway filled with liquid, then a small amount of the more volatile species ($N_2$ or CO) is injected in with the alkane. After sample mixing has been completed, the temperature is lowered in increments of 0.2–0.5 K from the initial equilibrium until the outburst is observed. Typically, the burst occurs soon after the first solid appears, where the temperature is then held until the pressure stabilizes. Once settled, the pressure is noted, and the cooling process resumes to observe how the temperature and pressure evolve over time.

Raman spectra are collected throughout the experiments, using a spectrometer that records shifts in the range of 100–3425 $cm^{-1}$, relative to the 875 nm laser wavelength. In this study, the spectra are expressly used to derive sample compositions. More precisely, we select the liquid spectrum taken directly before an outburst event and calculate the integrated band areas of each species in the mixture, using the $N_2$ peak at 2327 $cm^{-1}$, the CO peak at 2137 $cm^{-1}$, the $CH_4$ peak at 2905 $cm^{-1}$, and a compilation of five $C_2H_6$ peaks between 2850–3000 $cm^{-1}$ (Figure 3). While the CO, $N_2$, and lone $CH_4$ peaks can be fit with Gaussian profiles to a good approximation, the $C_2H_6$ peaks must be fit with Voigt profiles, which is a blending of Gaussian and Lorentzian profiles. Mole fractions corresponding to each peak fit are listed in the top right of the panels on the right hand side.

It should be noted that while there are isolated $C_2H_6$ peaks at lower wavenumbers, we have found the summation of the five C–H peaks (2880 cm$^{-1}$, 2882 cm$^{-1}$, 2916 cm$^{-1}$, 2938 cm$^{-1}$, and 2957 cm$^{-1}$) provide more accurate sample compositions for the $N_2/C_2H_6$ mixtures. We have confidence in the accuracy of our composition derivations, given our ability to estimate initial ratios from mixing using partial pressure, as well as having CRYOCHEM 2.0 equation of state (Tan & Kargel, 2018) as an external basis of comparison.

Several tests were conducted to determine the compositional uncertainty. The first tested the effect of spectral noise on derived compositions. For this test, we added artificial noise to a spectrum, derived the composition, and repeated the process, applying different noise values with each cycle. The overall effect of spectral noise on derived compositions was negligible. We determined that a larger contributor to compositional uncertainty came from the Gaussian and Voigt fits used to calculate the integrated band areas, as described earlier. The largest uncertainty in fit parameters to the data is ~1.2%, with the rest falling below this value. To encompass these sources of error, and account for any additional systematic errors and small instrumental effects, we have rounded our uncertainty for the derived compositions to be no larger than ±2%.

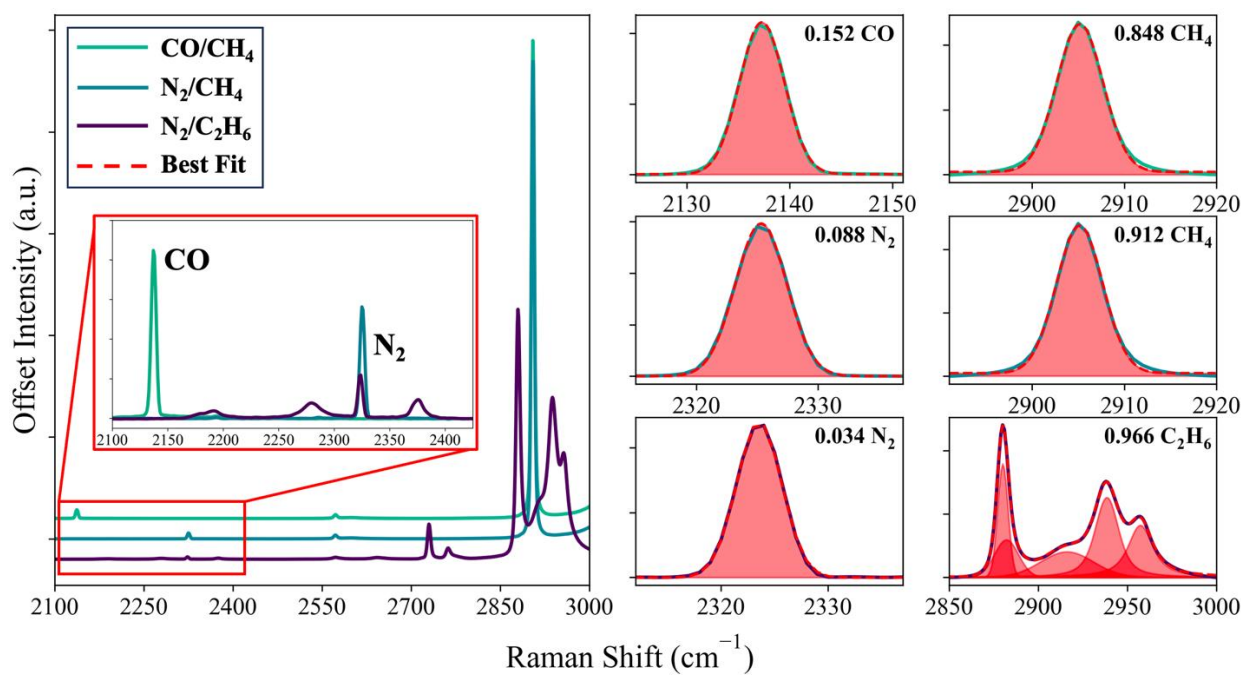

*Figure 3. Examples of Raman spectra with corresponding peak fits that are used for deriving compositions of the CO/CH$_4$, N$_2$/CH$_4$, and N$_2$/C$_2$H$_6$ binary systems. The derived compositions are shown for each peak fit in the top right of the right hand panels. The left panel shows the spectra between 2100–3000 cm$^{-1}$, where the peaks of interest are located. A zoomed in view of the comparatively weak CO (2137 cm$^{-1}$) and N$_2$ (2327 cm$^{-1}$) peaks is also provided. Note: the peaks at ~2275cm$^{-1}$ and 2375 cm$^{-1}$ in the N$_2$/C$_2$H$_6$ spectrum are from C$_2$H$_6$.*

## 3. RESULTS

The outbursts in the CO/CH$_4$, N$_2$/CH$_4$, and N$_2$/C$_2$H$_6$ binary systems occurred on cooling and began either at or slightly after freezing at the three-phase solid-liquid-vapor SLV boundary. Prior to freezing, the system is in two-phase liquid-vapor LV equilibrium. When freezing happens, solid phase emerges, so that the system is consequently in three-phase SLV equilibrium. On further cooling, the system stays in SLV equilibrium as the liquid transforms, and the whole system enters two-phase solid-vapor SV equilibrium after all liquid is gone. Therefore, when the burst happens, the system is still in the SLV equilibrium.

The onset of the events was partly evidenced by pressure spikes that occurred while lowering the temperature. The combination of increasing pressure with decreasing temperature is the defining characteristic of the negative slope on an SLV curve and is present between the three-phase maximum pressure point and the triple points of the less volatile species (values provided in Appendix B). Since the negative slope is positioned on the alkane-rich side of the N$_2$/CH$_4$, CO/CH$_4$, and N$_2$/C$_2$H$_6$ binary systems, it means the solid will be predominantly CH$_4$ or C$_2$H$_6$ once the SLV boundary is crossed. We were also able to verify outburst events visually through the emergence of distinct ice formations at the surface of the sample. Timelapses and pictures of the cell during these instances indicate the ice mounds form due to the release of the volatile-rich gas, which pushes the slurry mixture upwards in the process (timelapses are provided in the repository, Raposa & Engle, 2024). Through comparing the T–P data and videos, we found that the size of the structures that formed at the surface correlated to how much the pressure rose when the outburst event began. Here, we show how SLV curves may be predictors for outburst events in binary systems by comparing experiments with the CO/CH$_4$, N$_2$/CH$_4$, and N$_2$/C$_2$H$_6$ systems to previous work by Raposa et al. (2022). It is evident from the results that pressure, temperature, and composition all play a vital role in this freezing-induced outburst phenomenon.

Figure 4 shows an example of a pressure and temperature history from an outburst event in a 0.115 CO / 0.885 CH$_4$ mixture. Prior to the outburst, the pressure remained steady. Shortly after 12:00 (clock time) at 83.6 K, the pressure of the trapped fluid became sufficient enough to break through the confining ice, leading to an outburst and a sharp rise in pressure. After roughly 4 minutes, the pressure stabilized, though to ensure stabilization, the sample was held at the burst temperature for ~30 minutes. The pink stars at the bottom of the plot indicate where the pictures below were taken. Pictures A and B show that ice formation behaved as expected while pressure remained steady. However, shortly after this specific outburst, the escaping gas and liquid inflated into a mushroom-shaped ice structure, as shown in picture C. Over time, and after cooling the sample further, the height of the outburst material decreased, and the sample appeared to settle (picture D).

The pressure and temperature data, referred to here as cooling paths, collected during each set of binary experiments were overlaid onto their respective SLV curves reported by Raposa et al. (2022). Figure 5 shows the outbursts captured in the CO/CH$_4$ system at compositions of 0.059, 0.070, 0.115, and 0.152 CO mole fraction, with accompanying pictures of each sample post-burst. The 0.115 and 0.152 CO cooling paths follow closely along the SLV curve, which indicates that these samples retained the collective solid, liquid, and vapor phases for a longer time during the cooling sequence. In contrast, the 0.059 and 0.070 CO cooling paths diverged from the SLV curve, a consequence of quickly losing the liquid phase and being pushed into the solid-vapor regime.

The $N_2$/$CH_4$ system (Figure 6) behaves similarly to the CO/$CH_4$ system, with the 0.124 $N_2$ / 0.876 $CH_4$ cooling path following the trend of the SLV curve and the 0.035 $N_2$ / 0.965 $CH_4$ cooling path remaining mostly in the solid-vapor region. One slight deviation is in the 0.088 $N_2$ / 0.912 $CH_4$ sample, where the cooling path travels through the solid-vapor region until ~75 K, where it rejoins the SLV curve. Tables A1 and A2 in the Appendix show that outbursts happened closely after the freezing points for all $CH_4$ mixtures. Comparatively, the negative slope of the SLV curve (dP/dT) of the β-type $N_2$/$C_2H_6$ system (Figure 7) is much steeper than those of the CO/$CH_4$ and $N_2$/$CH_4$ systems; furthermore the cooling paths of the three $N_2$/$C_2H_6$ samples (0.013 $N_2$ / 0.987 $C_2H_6$; 0.024 $N_2$ / 0.976 $C_2H_6$; and 0.034 $N_2$ / 0.966 $C_2H_6$) traveled through the solid-vapor SV region at a nearly constant pressure until joining up the left side of the SLV curve.

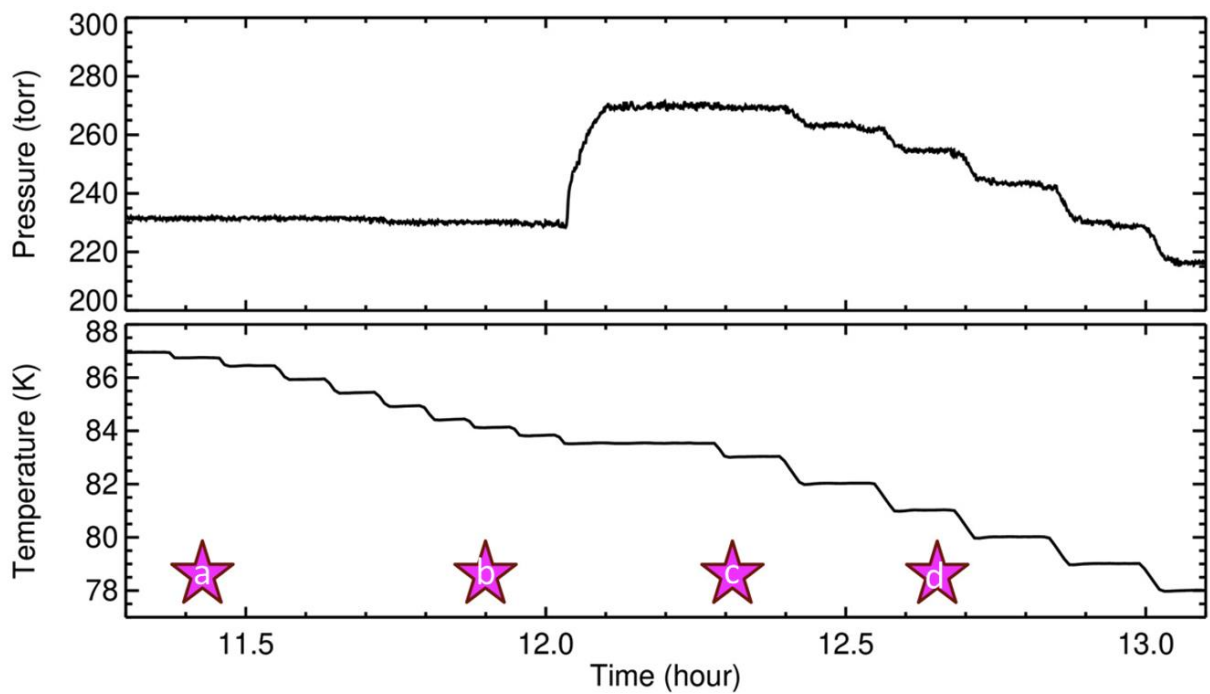
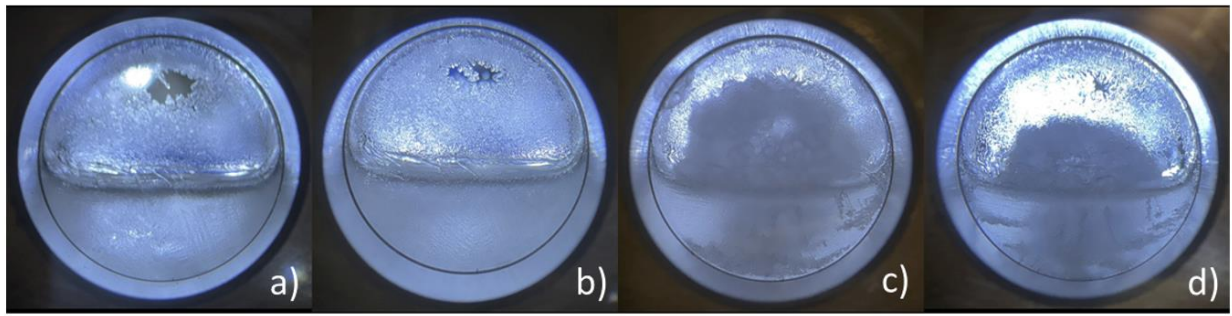

*Figure 4. Pressure and temperature history of a 0.115 CO / 0.885 $CH_4$ mixture. The pink stars indicate where the four photos were taken during the cooling sequence. The outburst occurred just after the 12-hour mark (clock time).*

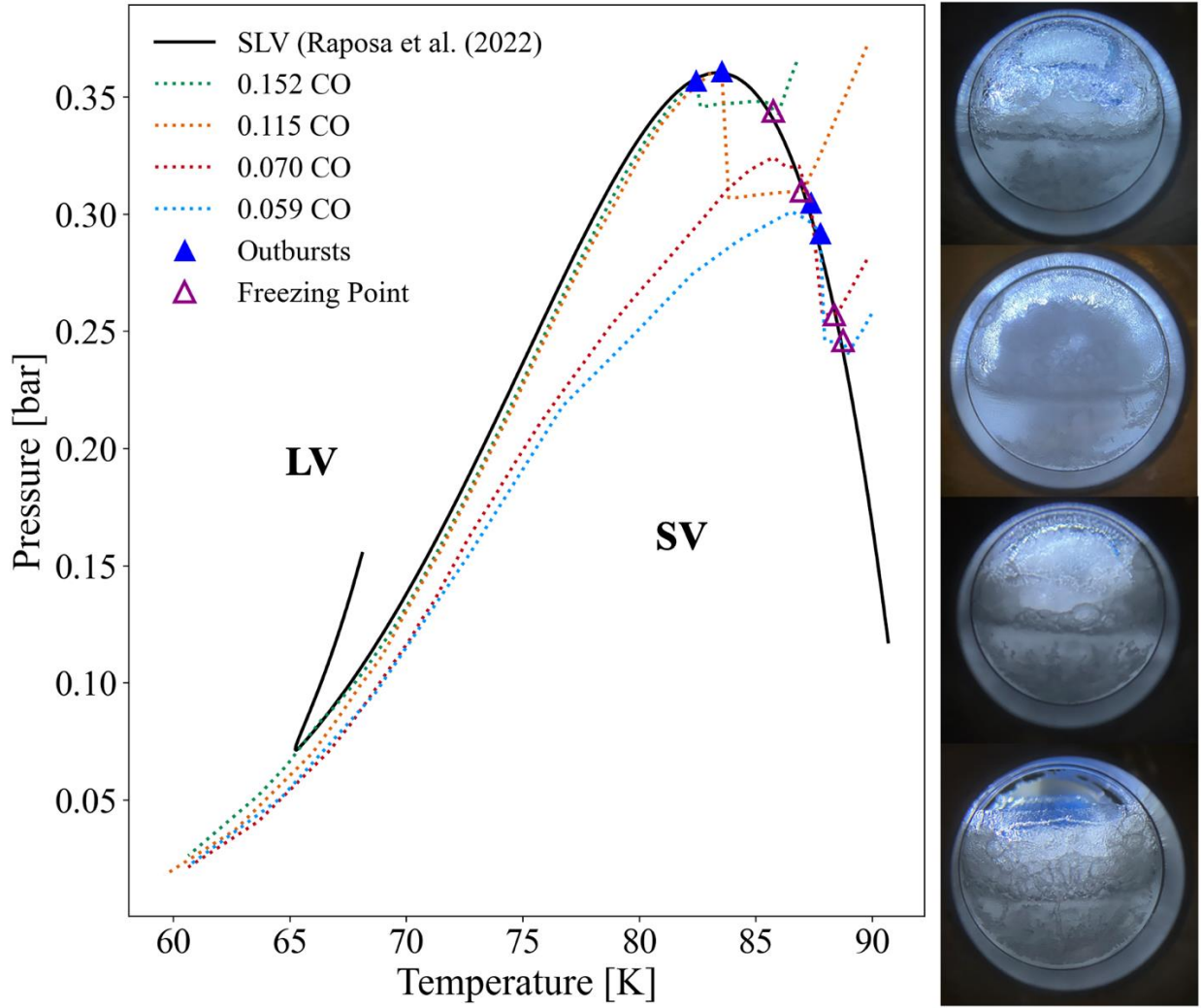

**Figure 5.** *Cooling paths for four different CO/CH$_4$ mixtures where outbursts occurred. CO mole fractions were derived just before the freezing point, when the sample was in the liquid phase. Pictures were taken right after the burst with mole fractions of CO from top to bottom: 0.152, 0.115, 0.070, and 0.059.*

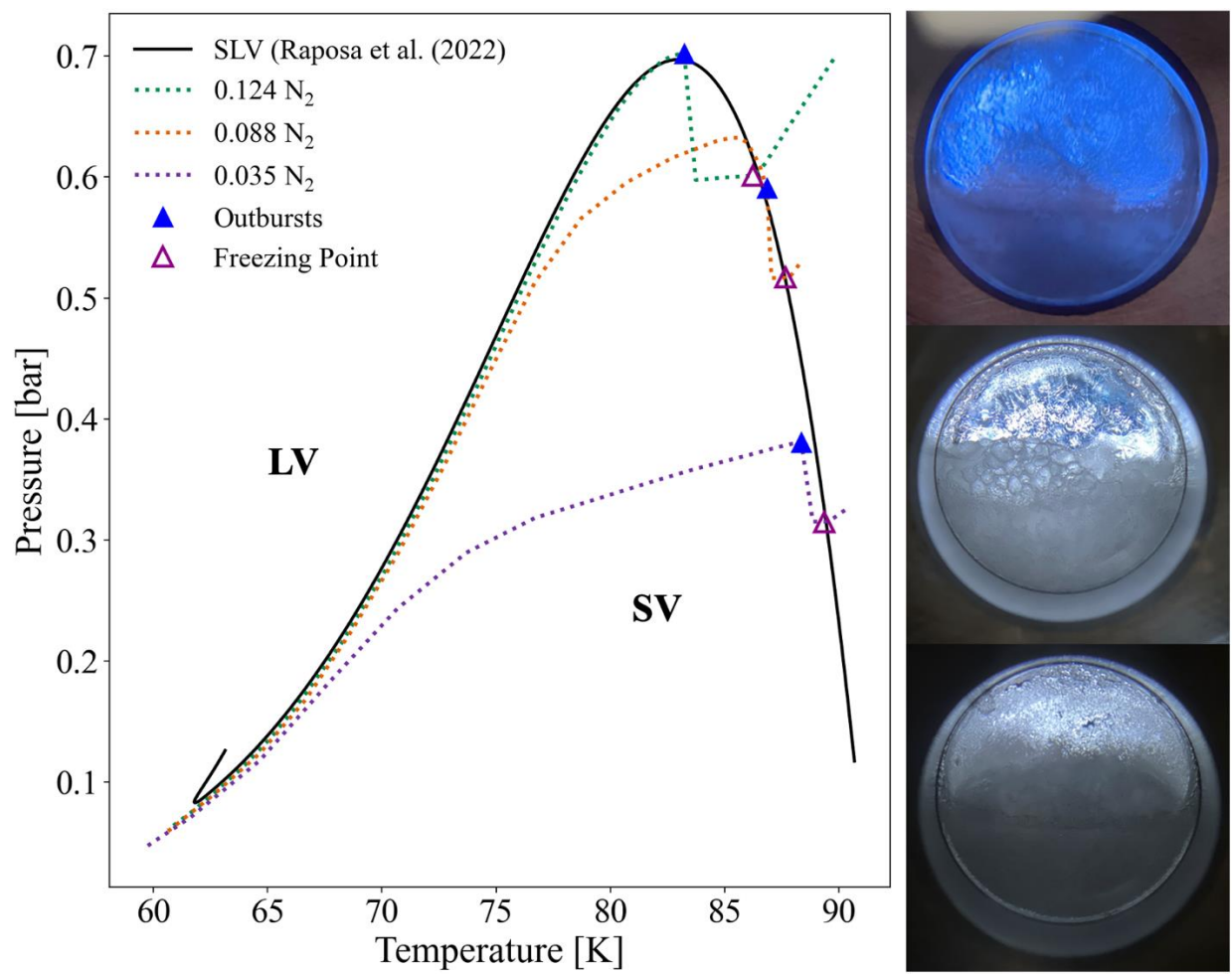

*Figure 6.* Cooling paths for three different $N_2/CH_4$ mixtures where outbursts occurred. $N_2$ mole fractions were derived just before the freezing point, when the sample was in the liquid phase. Pictures were taken right after the burst with mole fractions of $N_2$ from top to bottom: 0.124, 0.088, and 0.035.

There are a few additional similarities and differences to note between the $CO/CH_4$, $N_2/CH_4$, and $N_2/C_2H_6$ binary systems. While the outbursts occurred on the alkane-rich side for all three systems, some differences in composition should be noted. For the α-type SLV, $N_2/CH_4$ and $CO/CH_4$, the outbursts occurred for a larger range of compositions (up to ≲16% CO, and ≲13% $N_2$ initial liquid compositions). Additionally, the spectra indicate that the abundance of $N_2$ and CO pre- and post-burst were roughly unchanged in the condensed sample (Figure S1 in repository, Raposa & Engle, 2024). However, for the β-type SLV $N_2/C_2H_6$ system, the threshold for the amount of $N_2$ seen in the outbursts was <5% for pressures less than 1 bar and, in contrast to the α-type SLV curve, spectra show that all or nearly all $N_2$ outgassed following the outburst (Figure S2 in repository, Raposa & Engle, 2024). The general progression of freezing to outburst were also dissimilar between the two types. As depicted in Figures 5 and 6, the burst event of α-type SLV systems would occur abruptly well after initial freezing. However, the β-type SLV system would typically form ice and outgas at the same temperature, although with a small time delay; ice would form approximately 5 minutes into holding the temperature steady and would then undergo an outburst about 15-20 minutes after. This difference in outburst locations may be in part due to a supercooling effect exhibited by the β-type SLV $N_2/C_2H_6$ system, which is evident in Figure 7. Despite differences in the three binary systems, we found that as long as the composition of the mixture at freezing has an increasing pressure upon cooling, an outburst will occur.

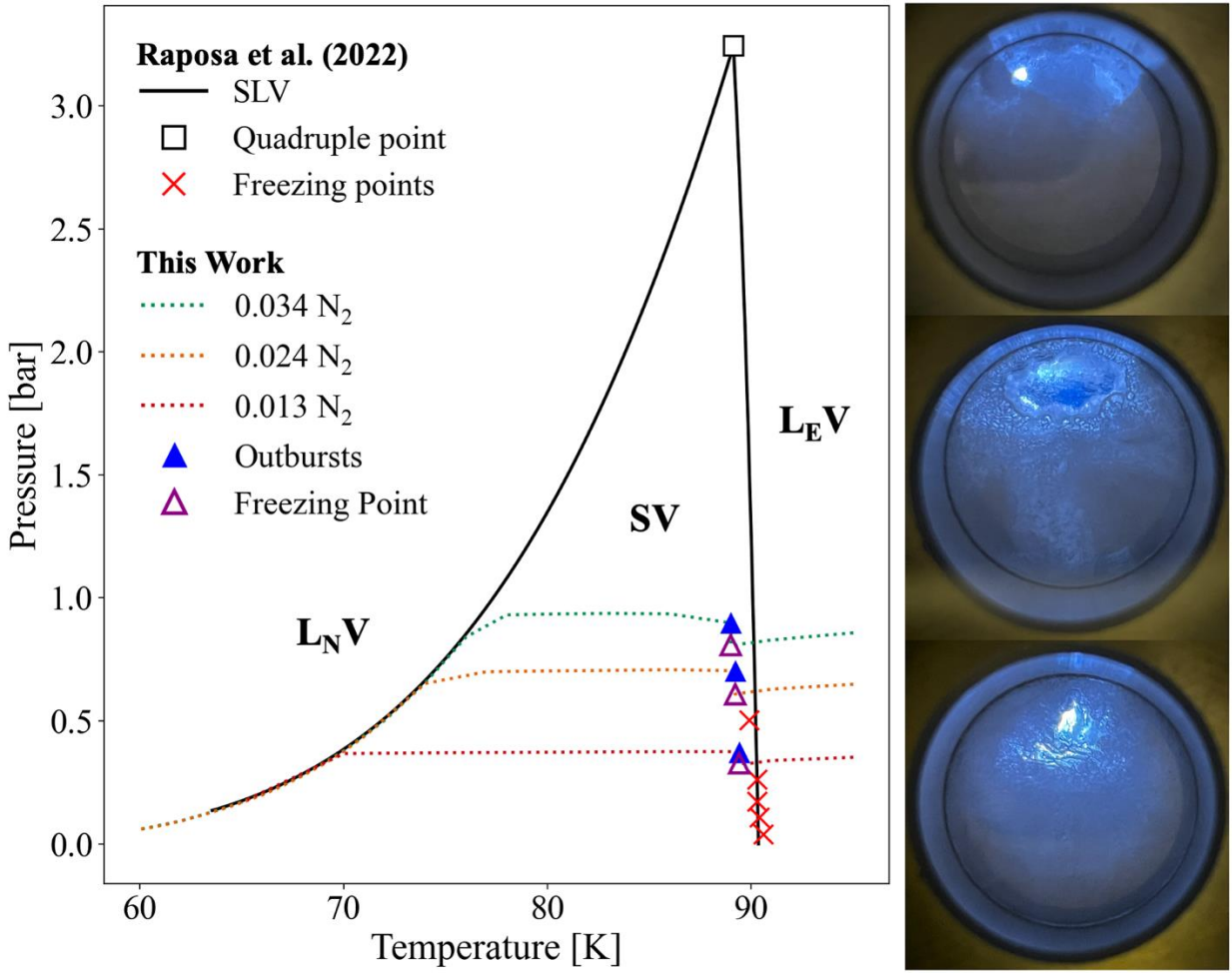

*Figure 7.* Cooling paths for three different $N_2/C_2H_6$ mixtures where outbursts occurred. $N_2$ mole fractions were derived at the temperatures shortly before the outbursts occurred (from top to bottom: 0.034, 0.024, 0.013) and the corresponding pictures were taken after each sample had stabilized post-burst. Subscripts represent phases that are more enriched in one species, where E=Ethane-rich and N=Nitrogen-rich.

Occasionally, the $N_2/C_2H_6$ samples would experience cycles of inflation and deflation before settling into a final surface morphology (Figure 8). When this occurred, the ice lid would expand upward, suddenly settle back to its initial height, and then repeat the process until the crust eventually froze in place. The movement was reminiscent of a breathing chest, although it should be noted that the ice lid grew due to inflation and not inhalation. The 'breathing' was presumably caused by an increase in pressure from gas build-up, which was released once the pressure was sufficient to overcome the strength of the lid. Of additional note was the apparent elasticity of the icy mound. There were no visible signs of cracking nor any instances of explosive outgassing, but only a soft deflation.

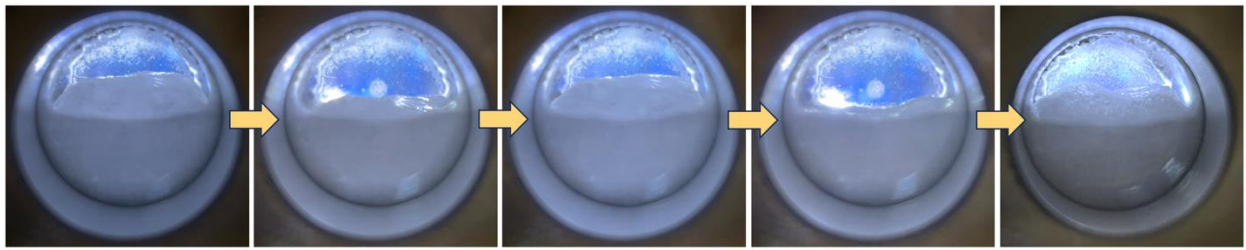

**Figure 8.** *An example of inflation/deflation cycles in a sample composed of 0.019 $N_2$ / 0.981 $C_2H_6$. The surface ballooned upward, settled back to the original level after outgassing, and repeated the process about 5 times. The furthest-right panel shows the final surface morphology.*

The cycles were also recorded through small pressure spikes shortly after the surface feature settled to its initial state. Both the pressure in the cell and the ice morphology would stabilize after approximately 30 minutes. A video showing an example of the 'breathing' can be found in the repository (Raposa & Engle, 2024). Interestingly, this phenomenon was not captured in either the CO/$CH_4$ or $N_2$/$CH_4$ systems, which may indicate it is a feature more characteristic of β-type SLV systems.

## 4. DISCUSSION AND IMPLICATIONS

These freezing-induced bursts require a mixture composition for which the pressure will increase on cooling as ice accumulates. This can occur when the mixture is rich in the less volatile component, corresponding to the part of the SLV curve with a negative slope, as seen in Figures 5–7. Additionally, the maximum pressure needs to exceed the atmospheric or confining pressure to be able to modify the environment. Here, we look at a few planetary settings where the outburst phenomenon might occur and extend our considerations to binary systems beyond those presented in Section 3, but which meet the criteria necessary for outburst events.

### 4.1. Subsurface Volatile Fluids on Pluto and Eris

The three main volatile species observed in Pluto's surface environment are $N_2$, CO, and $CH_4$. As shown in Section 3, $CH_4$ mixtures with either CO (Figure 5) or $N_2$ (Figure 6) exhibit pressure maxima along their SLV curves, reaching 0.7 and 0.35 bar, respectively (Raposa et al., 2022). A pressure of 0.7 bar corresponds to the overburden pressure at a depth of 120 m in a water ice crust on Pluto, or 220 m in less dense $CH_4$ ice (Bol'shutkin et al., 1971). The abundance of Pluto's volatile species prompts the question of whether shallow subsurface $N_2$/$CH_4$ or CO/$CH_4$ mixed fluids might enable eruptive processes as a result of gas release on freezing.

For this scenario to work, a $CH_4$-rich fluid is required. Several regions on Pluto host massive $CH_4$ ice deposits, probably with small amounts of $N_2$ and CO. These include the north polar region of Lowell Regio (Schmitt et al., 2017) and the bladed terrains of Tartarus Dorsa and apparently similar features on the sub-Charon hemisphere (Moore et al., 2018). With $CH_4$ ice acting as an insulating cap, the base of such a deposit may reach temperatures where partial melting could occur. Whether melting can happen within a few hundred meters of Pluto's surface depends on the internal heat flow (~0.003 W m$^{-2}$ on average, e.g., McKinnon et al. (2016), though it could be locally higher) and thermal conductivity of the CH$_4$ ice deposits (~0.3 W m$^{-1}$ K$^{-1}$ for a solid slab of CH$_4$ ice at 25 K; Jez-dotowski et al. (1997)). These numbers yield a thermal gradient of 10 K km$^{-1}$, placing the ~80 K melting point of a CH$_4$-rich ice mixture several km below the surface, but it could be much shallower if the CH$_4$ deposits on top are porous and granular since such a texture would be much more insulating. The initial melt from a N$_2$/CH$_4$ ice would be more N$_2$-rich than the solid. Depending on the relative abundances of CH$_4$ and N$_2$, the melt can be denser than the solid, and thus tend to descend (Raposa et al., 2022; Roe & Grundy, 2012). Although it would be less dense than Pluto's H$_2$O ice crust, it could still descend through cracks into an H$_2$O ice regolith, toward even warmer temperatures at greater depth. Such draining from the base of a CH$_4$ ice sheet might lead to formation of collapse pits, such as the clusters of pits north and east of Sputnik Planitia (Howard et al., 2017). After spending time at depth, the fluid could also be geothermally heated and driven upward. It would cool as it rises, and its eventual re-freezing would release N$_2$-rich gas that might trigger eruptive phenomena.

Eris' higher rock abundance and thus higher internal heat flow (~0.005 W m$^{-2}$), along with its high CH$_4$ ice abundance may provide even more favorable circumstances for this scenario. Its CH$_4$-rich surface deposits also contain relatively little N$_2$ (e.g., Grundy et al., 2024a).

### 4.2. Gas Emission Craters (GECs) on Earth

Interestingly, the findings from our study may offer a clue about the origin of high pressures at shallow subsurfaces in the formation of gas emission craters (GECs) in Siberia, Russia. As pointed out in the seminal paper by Buldovicz et al. (2018), GECs may have resulted from cryogenic phenomena rather than warming processes. They hypothesized that the explosion was triggered by a water-to-ice transition that increases the pressure, thus opposite to many other hypotheses that include gas migration from the deep due to global warming (Leibman et al., 2014). The phenomenon is then referred to as cryovolcanism on Earth in comparison to that found at the surface of icy bodies in the outer Solar System (Geissler, 2015).

As an alternative to the above hypotheses, the high pressure may be ascribed to the SLV of gas mixtures, in this case the binary of CH$_4$/CO$_2$, where the profile of the SLV curve on the P-T phase diagram is of α type as shown in Figure 1, with a huge negative slope prior to maximum pressure upon cooling. These two species are in fact major components of the gas measured around the Yamal crater found in 2014, in addition to N$_2$ and O$_2$ (Vorobyev et al., 2019) as being exposed to the open air after the explosion. Prior to the explosion, the original gas confined underground seemed to be dominated by CO$_2$ followed by CH$_4$, so that the SLV curve of the binary CH$_4$/CO$_2$ may be assumed to represent the whole gas. According to experimental data in the literature (Davis et al., 1962), the maximum pressure of the SLV is at 205 K and 48.7 bar, compared to the triple point of CO$_2$ (216.6 K and 5.19 bar). Therefore, cooling by 11.6 K from the triple-point temperature can increase the pressure by more than 43 bar if all phases are present in equilibrium, i.e., by a rate of 3.75 bar/K.

Comparing these data with the calculations from Buldovicz et al. (2018), which estimated the equilibrium pressure of the talik core under a 6-8 m thick cap is about 5 bar while about 10 bar is required to break down the cap, the additional 5 bar can be acquired by further cooling of just less than 1.5 K if the temperature of the core can reach ~216 K (~ –57 ºC). These temperatures are all within possibilities to happen in the area (e.g., the temperature in Yamal Peninsula may hit –60 ºC in winter (*Yamal Peninsula Travel: Seasons*, 2020).

When the subsurface gas is colder than the air above the surface, the large pressure difference due to the high pressure of the SLV of the confined gas underground can break the 8 m thick cap by explosion, thus creating the crater. Note that in this scenario, the necessary high pressure naturally arises from the SLV equilibrium of the binary gas upon cooling without invoking other mechanisms like gas-hydrate decomposition, which is only possible at low pressures; gas hydrates are stable at high pressures underground. Even if other mechanisms are considered, such as freezing water and volume reduction (Istomin et al., 2020), the SLV equilibrium may still potentially give a significant contribution. Nonetheless, this alternative origin of high pressure in the formation of GECs needs further evaluation and verification.

### 4.3. Triton South Polar Plumes

$CO_2$ and $CH_4$ have also been observed to be abundant as ices on the surface of Triton (Cruikshank et al., 1993; Quirico et al., 1999), Neptune's large, captured moon. The lack of longitudinal variability in Triton's $CO_2$ absorption bands as Triton rotates on its axis (Grundy et al., 2010; Holler et al., 2016) has been taken as evidence that the $CO_2$ ice is located around the south polar region, the same region where several active eruptions were observed during the Voyager 2 flyby in 1989 (Hansen et al., 1990; Smith et al., 1989; Soderblom et al., 1990). The eruptions produced plumes that were observed to rise vertically to an altitude of about 8 km before extending downwind as far as 150 km.

Several hypotheses have been proposed to account for Triton's plumes, as reviewed by Hofgartner et al. (2022). We offer another, based on the mechanism explored in this paper. In our scenario, the eruptions on Triton are another example of explosive gas emissions instigated by freezing-induced overpressurization, involving a mixture of $CO_2$ and $CH_4$. The $CO_2$ could be sourced from Triton's interior ocean. The $CH_4$ could share that source or could be diffusing out of a clathrate layer at the interface between ocean and crust (e.g., Howard et al., 2023). It could also come from near-surface $CH_4$ ice deposits, considering that $CH_4$ is abundant on the surface and mobile at Triton's surface temperatures (e.g., Grundy et al., 2024b; Quirico et al., 1999). As a $CO_2$ and $CH_4$ mixture rises through Triton's crust, it will cool, leading to the preferential freeze out of $CO_2$ ice while $CH_4$-rich gas is produced, leading to a large pressure increase. The strong pressure rise could enable solid particles of $CO_2$ to be entrained in the flow, perhaps accompanied by particles of $H_2O$ ice or other solids from Triton's crust. Warm $CH_4$ gas would be buoyant in Triton's $N_2$-dominated atmosphere, helping to explain the initially near vertical rise of the plumes. Small $CO_2$ ice particles would not sublimate away at the ~40 K temperatures of Triton's surface environment, which may account for the long spatial extent of the visible plume clouds.

### 4.4. Sharp-Edge Depressions (SEDs) on Titan

Titan has become well known for its polar lakes and seas and for the mysteries surrounding their formation mechanisms. The Cassini mission showed many of the lacustrine features to be wide and lined with sharp-edge depressions (SEDs) (Birch et al., 2019; Solomonidou et al., 2020). Additionally, many of the small lake basins were found to have raised rims and are either circular or irregular in shape (Birch et al., 2019). Similar to the work by Buldovicz et al. (2018) concerning the Yamal Peninsula GEC, Mitri et al. (2019) proposed the SEDs may have formed through explosive mechanisms, although their assumption was based on gas release as a consequence of heating, not freezing. Nonetheless, there are many similarities between the SEDs on Titan and the Yamal GEC, raising the question of whether the smaller lake basins on Titan could have also formed as a consequence of freezing-induced overpressurization. Of note, Mitri et al. (2019) indicates that 1) the explosions would likely have some dependence on a phase transition of a volatile material; 2) both studies suppose the presence of a thermokarst preceding the explosive events; 3) there are tephra-like features with dissimilar compositions to the basins surrounding the SEDs, suggesting excavation; and 4) the circular shape and raised rims are morphologically similar to that of the Yamal GEC.

However, there is a noticeable difference between the dimensions of the SEDs and GECs. The Yamal GEC is reported as being 20 m in diameter and ~52 m deep (Buldovicz et al., 2018); in contrast, a typical SED is approximately 100 km across and 200 m deep (Birch et al., 2019), with some of them being non-circular. This discrepancy may be attributed to a collection of factors. In regard to the periodic irregularity in shape, it was proposed it may be due to multiple closely spaced explosive vents, which is a feature not uncommon in the formation process of terrestrial maars (Mitri et al., 2019;). As for the difference in width, it is likely that smaller SEDs exist, possibly on the scale of the Yamal GEC, but the low spatial and vertical resolutions of the Cassini topographic data meant they could not be detected (Birch et al., 2019). Furthermore, the resulting geologic features created through overpressurization are informed by properties such as surface gravity, tensile strength of the substrate, composition of the substrate and accompanying volatiles, and atmospheric pressure, meaning different planetary environments will have explosive features of differing dimensions (P. Brož et al., 2023; Holsapple & Schmidt, 1980). This is exemplified by Mitri et al. (2019) through comparisons of tensile strength between crustal materials on Titan and on Earth. Experiments have found the tensile strength for water ice on Titan to be between 0.1–2.2 MPa (Litwin et al., 2012), whereas basaltic rock on Earth is closer to 15 MPa. These results indicate explosions could be deeper and form out of smaller quantities of volatile material on Titan as those occurring on Earth (Goto & Taniguchi, 2001; Holsapple & Schmidt, 1980; Mitri et al., 2019).

While the comparisons between Earth GECs and Titan SEDs are compelling, there are other aspects to consider for freezing-induced cratering. Mitri et al. 2019 propose that the explosions started at an earlier time in Titan's history when it may have had surface temperatures as low as 81 K (Lorenz et al., 1997). At this temperature, large pools of liquid $N_2$ could have existed on the polar surface, with some of it likely infiltrating into the subsurface. As Titan warmed, overpressurization of the trapped $N_2$ could have instigated an explosive eruption, leaving behind a crater. Given that outbursts can occur in the $N_2/C_2H_6$ system, it may be that SEDs have been forming throughout Titan's history as a result of both heating and freezing overpressurization mechanisms. Although the experiments presented here only pertain to pressures at <1 bar, the calculated SLV curve indicates outbursts might be possible up to ~3.25 bar and between a narrow temperature range of 89–90 K (Figure 7). One possible scenario for $N_2/C_2H_6$-driven explosive events may be the freezing of $C_2H_6$-rich subsurface reservoirs, with infilled karstic caves (Malaska et al., 2022; Wynne et al., 2022) or pockets formed through the upwelling of gas that has dissociated from clathrate hydrates (Fagents et al., 2022) acting as the reservoirs beneath Titan's crust. Once in the cavity, if a sufficient amount of $N_2$ were to be introduced, it could instigate ice forming in the $C_2H_6$-rich liquid, which would lead to overpressurization followed by an outburst. It should be noted that this process is likely not feasible directly at the surface at the present conditions at Titan's poles (~89 K, 1.5 bar). Namely, because the lower temperatures allow for $CH_4$-rich liquid to pool, which hinders the outburst mechanism (Mastrogiuseppe et al., 2018, 2019; Tan et al., 2015). However, it may be that the $C_2H_6$-rich subsurface reservoirs underwent the explosive events at an earlier time and the resulting lake basins were later filled with the $CH_4$-rich liquid seen today.

### 4.5. Ice-Mud Volcanism on Mars

On Earth, volcanism can be broken down into two main categories: igneous and sedimentary (Mazzini & Etiope, 2017). Both volcanic systems are gas-driven and involve subsurface processes that end in fracturing and release of material onto the surface (Mazzini, 2015). However, as the names suggest, the two categories differ in composition, thus leading to a contrast in resulting features. Mud (i.e., sedimentary) volcanism is propelled by volatile liquids and gasses that are often composed of water, saline water, $CO_2$, $N_2$, $CH_4$, $C_2H_6$, and larger hydrocarbons (Mazzini & Etiopa, 2017); materials that are prevalent on both terrestrial and icy worlds in the solar system.

Experiments carried out by Brož et al. (2023) produced strikingly similar results to those captured in the Astrophysical Materials Lab, albeit in the context of terrestrial materials subjected to Martian conditions. The study investigated the volumetric changes associated with mud-silicate mixtures of three different viscosities when exposed to a pressure drop from 1 bar down to ~6 mbar and between 273–300 K. These experiments were prompted by previous work suggesting the instability of water at Martian conditions could result in outgassing (Bargery et al., 2010; Wilson & Mouginis-Mark, 2014). Brož et al. (2023) not only confirmed this but also found the low pressure triggered evaporative cooling of the mud, leading to the formation of a frozen crust that obstructed bubbles from immediate escape. Depending on the viscosity of the mud and the selected conditions, this would result in a variety of end products, to include the formation of prominent domes, smaller propagating lobes, dome collapse, ejection of material, and cycles of inflation/deflation that resembled breathing. We can confidently report dome formation and dome collapse in our binary systems (and inflation/deflation cycling in the $N_2/C_2H_6$ system), although it is uncertain whether fluid propagation or material ejection is possible. However, there may be hints that flow could occur.

In the Martian experiments, samples would occasionally exhibit vertical growth due to the container size, which would result in piling of material and eventual lateral spread. We similarly saw upward expansion of material that appeared to be confined by the size—and possibly shape—of the sample cell. The upward expansion was characterized by liquid continuing to pool at the surface as the underlying material froze, which increased the sample volume and sometimes completely overfilled the cell. The vertical growth and continued presence of surface liquid may indicate that certain conditions could induce flow if provided sufficient space (see repository for timelapse, Raposa & Engle, 2024). This may tie into another study simulating Martian mud-lava flows (Brož et al., 2020). The work showed that as the flow propagated, the mud would freeze both at the surface and at its base, consequently overpressurizing the residual interior liquid, creating voids and lobes as it spreads. If our icy systems can in fact produce flow, the behavior may be similar to that of the mud-lava flows seen by Brož et al. (2020).

Of course, the mud-silicate experiments have an apparent contrast to those of the three hydrocarbon-rich binary systems we investigated. Most notably, the volatile liquid used by Brož et al. (2023) is only composed of water and does not include a second, more volatile species. While this initially makes for a more tenuous connection, the study is still poignant to our discussion of freezing-induced outbursts. Specifically, the mud mixture may be a terrestrial analog to the icy slurries formed in the hydrocarbon-rich systems; the solid material is still the less 'volatile' species and the trapped volatile liquid and gas are the cause of the overpressurization. It was also shown that the 'low' viscosity sample, composed of 75 wt% water / 25 wt% clay, did not overpressurize and thus did not result in an outburst. The outcome aligns with our observations and predictions that mixtures with high volatile concentrations do not experience an outburst. And, while the mud-silicate mixture may not have an SLV curve, this may simply mean there are other material properties that can be used as a predictive measure.

It would be of additional interest to see a study carried out with ammonia ($NH_3$) as the secondary volatile species. The SLV curve for the $NH_3/H_2O$ system also fits the α-type profile and reaches a maximum pressure of 18.12 mbar (Lugo et al., 2006), which is ~3x higher than Mars' surface pressure. Reviews of IR and MIR spectra indicate the presence of $NH_3$ in Mars' atmosphere (Trokhimovskiy et al., 2024; Villanueva et al., 2013), which is an important finding as $NH_3$ quickly photolyzes in the atmosphere, meaning there must be a replenishment source. It has been hypothesized that $NH_3$ could reside in the surface or subsurface (Mancinelli & Banin, 2003; Summers et al., 2012) and may be produced through reactions between nitric oxide (NO) and iron sulfide (FeS) (Summers et al., 2012). If $H_2O$ is also present below the surface, this could result in outbursts, especially if the $NH_3/H_2O$ mixture is ≤10% $NH_3$, as is implied by the SLV curve of the binary system (Lugo et al., 2006). Further, the negative slope of the SLV curve is between -18–0 ºC (~250–273 K), which is well within the expansive Martian surface temperature range of -143–20 ºC (~130–300 K) (Tillman et al., 2006).

Beyond Earth and Mars, studies are showing that mud volcanism may be a common geologic process on other bodies and may be partially driven by freezing-induced overpressurization. Additional work suggests mud volcanism, effusive flows, and their cryogenic analogs may be occurring on other planetary environments such as the Moon (Wilson & Head, 2017), Ceres (Neveu & Desch, 2015; Ruesch et al., 2019), Europa (Fagents, 2003; Quick et al., 2022), Enceladus (Brož et al., 2023), Titan (Fortes & Grindrod, 2006), Pluto (Ahrens, 2020; Martin & Binzel, 2021), and any number of TNOs and KBOs (Neveu et al., 2015).

## 5. CONCLUSIONS

While conducting experiments on the $CO/CH_4$, $N_2/CH_4$, and $N_2/C_2H_6$ systems, we witnessed outbursts of volatile materials from condensed samples upon freezing. These events were not only characterized by pressure spikes in the sample cell but also through surface features created by explosive and effusive processes. Although similarities in the systems were mostly seen between the two $CH_4$ binary systems, there was one notable commonality shared by all three systems. That is, the outbursts occurred along the negative slopes of the three SLV curves—where the pressure increases with decreasing temperature—between the triple point of the less volatile component and the three-phase maximum pressure point along the curve. This commonality indicates the negative slope is a necessary thermodynamic condition to allow the outbursts to occur on cooling.

Quinzio et al. (2023) recently designed a classification structure for SLV curve topologies of binary mixtures. We found that the $N_2/CH_4$ and $CO/CH_4$ systems fit the α-type profile (Figure 1), whereas the $N_2/C_2H_6$ system is more akin to the β-type (Figure 2), with the primary distinction being that the β-type system peaks at an SLLV quadruple point and has a much steeper negative slope as compared to the α-types. That being said, a few differences were noted between the α-type $N_2/CH_4$ and $CO/CH_4$ systems and the β-type $N_2/C_2H_6$ system. For the α-types, the outbursts occurred in a larger range of compositions, up to ≲16% CO and ≲13% $N_2$ initial liquid mole fractions. Conversely, the event only occurred in samples of <5% $N_2$ in the $N_2/C_2H_6$ system at <1 bar. Furthermore, spectra indicate that the abundances of $N_2$ and CO in the condensed samples were roughly unchanged pre- and post-burst, whereas the $N_2/C_2H_6$ condensed samples lost practically all of the $N_2$ to outgassing. We additionally found that outbursts happened at colder temperatures than those of the freezing points for the α-types (Figures 5 and 6), and saw that the bursts occurred at the same temperatures as the freezing points in the β-type system (Figure 7) but with a ~20 minute delay between the two events.

We also saw physical differences in the outburst behavior amongst all three systems, with either explosive or effusive ice morphologies being a possible end result. Sometimes, the outgassed material would cause the slurry to balloon up and form one large mushroom-shaped feature (ex. Figure 4, Picture C). Other times, the liquid level would slowly rise and solidify as the gas bubbled out. And occasionally, samples would undergo a cycle of inflation and deflation (Figure 8) before settling to a final state, although this behavior was only witnessed in the $N_2/C_2H_6$ system. It should also be noted that cratering may have also occurred at some point but was not documented because ice forming on the window obscured it from view.

Along with the three binary systems, preliminary experiments have indicated outgassing events occurring in the $N_2/CH_4/C_2H_6$ system, as well as the $N_2/C_2H_6/C_3H_8$ (example in repository, Raposa & Engle, 2024) and $N_2/CH_4/C_2H_6/C_3H_8$ systems. Currently, we have not seen any indications of surface roughening in these systems, but we anticipate a more rigorous examination of them in the future. It should also be noted that we have not seen signs of bursts in preliminary $N_2/CO$ experiments, or in the $CH_4/C_2H_6$ system (Engle et al., 2021). In the case of the $N_2/CO$ system, the lack of outburst occurrences may be due to the small magnitude of the maximum pressure, i.e., 0.159 bar, compared to the triple-point pressure of CO at 0.153 bar (Raposa et al., 2022). The same for the $CH_4/C_2H_6$ system, where the SLV curve shows a small maximum pressure of less than 24 mbar on the ethane-rich branch of the phase diagram (Wang et al., 2023). Given the results of the binary experiments presented here and the outgassing from the $N_2/CH_4/C_2H_6$ system that represents Titan's fluids, it would be worthwhile to test $CH_4$-rich $N_2/CH_4/CO$ mixtures that represent Pluto's surface/subsurface materials for burst behavior. Generally, investigating >2-component mixtures is an important next step in assessing how these outbursts might be shaping planetary bodies in the solar system and how we can more accurately predict the behavior in more complex systems.

The findings suggest similar processes may be occurring on both terrestrial and frozen worlds that uniquely provide sufficient conditions, i.e., for volatile materials being brought to the surface due to freezing-induced overpressurization on those individual bodies. On Pluto and Eris, explosive phenomena could occur in $CH_4$-rich $CH_4/N_2$ subsurface fluids below thick $CH_4$ ice deposits. On Earth, gas emission craters found in Siberia, Russia, could be made by outbursts of $CO_2$-rich $CO_2/CH_4$ gas that breaks a ~6-8 m thick cap. Similarly, outbursts of $CO_2$ from a $CO_2/CH_4$ subsurface mixture could explain plumes seen on Triton. A number of lakes and lake basins on Titan have been categorized as 'sharp-edge depressions' and may have formed by explosive processes, perhaps as a consequence of $C_2H_6$-rich $N_2/C_2H_6$ subsurface reservoirs. On Mars, overpressurization in $H_2O$ ice-mud slurries could be responsible for mud volcano-like surface features and effusive sedimentary-rich lava flows. And, if $NH_3/H_2O$ mixtures reside in the Martian subsurface, this could result in explosive events considering the planet's low surface pressure and expansive temperature range. In relation, processes similar to mud volcanism may also be occurring in the outer solar system, with slurries and liquids composed of volatile materials.

There are still a number of studies that can be performed that would provide a clearer understanding of the outgassing process and how it might apply to geophysical phenomena occurring on planetary settings. This may include: 1) Utilizing a facility that provides a larger sample surface area, which would aid in determining how sample volume affects burst behavior and whether liquid flow is possible. 2) Investigating other binary systems at low pressures such as $NH_3/H_2O$ (as described in Section 4.4), Ar/Kr (Heastie, 1959), and $Ar/CH_4$ (Van 'T Zelfde et al., 1968), which are relevant to the outer solar system and are known to have a negative slope along the SLV curve. 3) Using an apparatus that can operate at pressures higher than 3.5 bar (the upper limit for the Astrophysical Materials Lab) in order to investigate suspected outburst systems such as $CH_4/CO_2$ in Figure 1 (as described in Sections 4.2 and 4.3) and $CH_4/H_2S$ (Kohn & Kurata, 1958) in Figure 2. While not expanded on in this paper, the $N_2/CO_2$ system (Fandiño et al., 2015) should also be tested for outburst behavior, as it exhibits a very steep negative slope on the SLV curve, has an unknown maximum pressure that is >130 bar, and could have important implications for $CO_2$ jets on Mars (Portyankina et al., 2017).

A logical near-future direction is to use these experimental results in models of planetary surfaces and sub-surfaces. For example, we may treat the flow of such binary mixtures within the subsurface of bodies like Pluto and Titan. Such fluids could be circulating through a refractory porous medium (e.g., say through a porous $H_2O$ ice crust), being driven by externally applied pressures or perhaps even due to thermally driven buoyancy effects due to the heat flux emerging from the interior. Under these circumstances and given the fluid's composition, pressure/temperature fluctuations along the fluid's flow path can result in local conditions favoring deposition and/or evaporation, driving subsurface regional overpressurization as well as modifying the region's local porosity. These effects will prove to be especially consequential if the binary fluid is at or very near its triple point. Simplified models examining this dynamic would thus be instructive and offer physical insights.

## ACKNOWLEDGEMENTS

Portions of this work and laboratory facility were supported by NASA Solar System Workings Program grants 80NSSC21K0168, 80NSSC18K0203 and 80NSSC19K0556, as well as NASA FINESST award (NNH19ZDA001N-FINESST). We are also grateful for philanthropic support from the John and Maureen Hendricks Foundation and from Lowell Observatory's Slipher Society. We thank Kendall Koga and David Trilling for insightful scientific discussions that helped motivate ideas presented in this paper.

## DATA AVAILABILITY STATEMENT

Data and timelapses are stored on figshare (https://doi.org/10.6084/m9.figshare.c.7193250, Raposa & Engle, 2024).

# APPENDIX

## A. Thermodynamics of the SLV equilibrium of binary mixtures

For a binary mixture A/B with A as the more volatile component, the triple-point temperatures satisfy the inequality $T_A < T_B$. Upon cooling from $T_B$ to $T_A$, the three-phase solid-liquid-vapor (SLV) equilibrium represents a vapor-liquid equilibrium with an emerging solid, i.e., the experimental freezing point in this work, which obeys the Clapeyron equation (Equation A1):

$$dP/dT = (\Delta h / \Delta v)(1/T) \qquad \textbf{Eq. (A1)}$$

The derivative is the slope along the SLV curve on the P-T phase diagram. $\Delta h$ and $\Delta v$ are the change of molar enthalpy (also known as the latent heat of phase transition) and the corresponding change of molar volume, respectively, due to the phase transition(s) at the equilibrium.

It has been known for many binary mixtures that the slope of the SLV curve is initially negative upon cooling, then positive after reaching a maximum, as shown in Figure 1 for α-type SLV. As the cooling proceeds, liquid is transforming into vapor and solid to allow the maximum pressure. For a given amount of liquid that undergoes phase transitions at T in the SLV equilibrium, if the mole fraction of liquid transforming into solid is σ, then that into vapor is (1 – σ). The corresponding total heat of the phase transitions and total change of volumes per mol liquid (Equations A2 and A3):

$$\Delta h^{L \to S+V} = \sigma \, \Delta h^{LS} + (1 - \sigma) \, \Delta h^{LV} \qquad \textbf{Eq. (A2)}$$

$$\Delta v^{L \to S+V} = \sigma \, \Delta v^{LS} + (1 - \sigma) \, \Delta v^{LV} \qquad \textbf{Eq. (A3)}$$

where $\Delta h^{LS}$ is the latent heat of freezing per mol liquid ($\Delta h^{LS} < 0$), while $\Delta h^{LV}$ is the latent heat of evaporation per mol liquid ($\Delta h^{LV} > 0$). The change of molar volume $\Delta v^{LS}$ is negative for freezing and $\Delta v^{LV}$ is positive for evaporation, but it is well known that the molar volume change $|\Delta v^{LS}| \ll |\Delta v^{LV}|$, so that $\Delta v^{L \to S+V}$ may be assumed to be always positive. Therefore, the change of sign of the SLV slope is due to the latent heat resulting from the competition between freezing and evaporation. The pressure reaches maximum when the freezing releases heat at the same amount as that absorbed by the evaporation. In other words, the first term on the right-hand side of Eq. A2 has the same absolute value as the second term, thus both terms sum to zero. Prior to the maximum upon cooling from a liquid-vapor equilibrium, the total released heat of freezing is more than that needed for the accompanying evaporation, and the other way around after the maximum. For this flip of heat to happen, the mole fraction σ needs to decrease upon cooling, which means more vapor is formed. For β-type SLV systems with a quadruple point, as those shown in Figure 2, the above description is only valid for the part with negative slope before the quadruple point upon cooling; another mechanism of phase transitions takes place beyond the point as a new liquid phase replaces the old one. Therefore, the total latent heat is not zero at the quadruple point. Nonetheless, only the former part is relevant to this work.

An important consequence of this type of cooling process along the negative slope of the SLV curve is that vapor is also appreciably produced along with the formation of solid phase that together release a net amount of latent heat. Note that if the evaporation is negligible, i.e., $\sigma \approx 1$, both Eq. A2 and Eq. A3 are negative thus a monotonic positive slope in Eq. A1, which means the absence of maximum pressure. For cases where the resulting vapor is confined inside a closed volume, in addition to the increasing pressure upon cooling prior to the maximum pressure, the net released latent heat may be taken by the vapor to do some work to expand further by pushing the confinement. Depending on the material strength of the confining solid, the vapor, sometimes with the inclusion of its liquid, may escape violently as a burst.

**B. Experimental data for freezing and bursts**

Experimental data for the freezing and burst points for the CO/$CH_4$ (Figure 5), $N_2$/$CH_4$ (Figure 6), and $N_2$/$C_2H_6$ (Figure 7) systems, along with the differences between those values, can be found in Tables A1–A3. Table A4 lists the maximum pressure points as measured in previous work (Raposa et al., 2022) and triple points for $CH_4$ and $C_2H_6$ from NIST Chemistry WebBook. The triple points for the alkanes are provided since the outbursts occur on those sides of the SLV curves.

*Table A1. Experimental data of the freezing and burst points for the CO/$CH_4$ system. Values in parentheses belong to points of maximum pressures that were reached after the bursts.*

| CO/$CH_4$ | Freezing Point | | Burst Point | | Difference | |
|---|---|---|---|---|---|---|
| CO [mol frac] | $T^{SLV}$ [K] | $P^{SLV}$ [bar] | $T^{burst}$ [K] | $P^{burst}$ [bar] | $\Delta T$ [K] | $\Delta P$ [bar] |
| 0.059 | 88.76 | 0.246 | 87.76 | 0.292 | -1 | +0.046 |
| | | | (86.55) | (0.301) | (-2.21) | (+0.055) |
| 0.070 | 88.36 | 0.257 | 87.36 | 0.305 | -1 | +0.048 |
| | | | (85.75) | (0.324) | (-2.61) | (+0.067) |
| 0.115 | 86.96 | 0.310 | 83.54 | 0.361 | -3.42 | +0.051 |
| 0.152 | 85.75 | 0.344 | 82.43 | 0.357 | -3.32 | +0.013 |

***Table A2.*** *Experimental data of the freezing and burst points for the N₂/CH₄ system. Values in parentheses belong to points of maximum pressures that were reached after the bursts.*

| $N_2/CH_4$ | Freezing Point | | Burst Point | | Difference | |
|---|---|---|---|---|---|---|
| $N_2$ [mol frac] | $T^{SLV}$ [K] | $P^{SLV}$ [bar] | $T^{burst}$ [K] | $P^{burst}$ [bar] | $\Delta T$ [K] | $\Delta P$ [bar] |
| 0.035 | 89.37 | 0.315 | 88.36 | 0.381 | -1.01 | +0.066 |
| 0.088 | 87.66 | 0.517 | 86.86 (85.65) | 0.591 (0.633) | -0.8 (-2.01) | +0.074 (+0.116) |
| 0.124 | 86.25 | 0.601 | 83.24 | 0.702 | -3.01 | +0.101 |

***Table A3.*** *Experimental data of the freezing and burst points for the N₂/C₂H₆ system.*

| $N_2/C_2H_6$ | Freezing Point | | Burst Point | | Difference | |
|---|---|---|---|---|---|---|
| $N_2$ [mol frac] | $T^{SLV}$ [K] | $P^{SLV}$ [bar] | $T^{burst}$ [K] | $P^{burst}$ [bar] | $\Delta T$ [K] | $\Delta P$ [bar] |
| 0.013 | 89.40 | 0.328 | 89.40 | 0.371 | 0.0 | +0.043 |
| 0.024 | 89.20 | 0.609 | 89.20 | 0.703 | 0.0 | +0.094 |
| 0.034 | 89.00 | 0.807 | 89.00 | 0.902 | 0.0 | +0.095 |

*Table A4. Maximum pressure point values for the SLV curves of each binary mixture (Raposa et al., 2022) and alkane triple point values.*

| Binary Mixture | Maximum Pressure Point | |
|---|---|---|
| | T [K] | P [bar] |
| $CO/CH_4$ | 83.18 | 0.352 |
| $N_2/CH_4$ | 82.89 | 0.699 |
| $N_2/C_2H_6$ | 89.12 | 3.244** |

| Alkane | Triple Point* | |
|---|---|---|
| | T [K] | P [bar] |
| $CH_4$ | 90.67 | 0.1169 |
| $C_2H_6$ | 91.00 | $1.1 \times 10^{-5}$ |

*Quadruple point (calculated in Raposa et al., 2022)

**NIST Standard Reference Database 69: NIST Chemistry WebBook

# REFERENCES

Ahrens, C. J. (2020). Modeling cryogenic mud volcanism on Pluto. *Journal of Volcanology and Geothermal Research*, *406*, 107070.

Ahrens, Caitlin J., Grundy, W. M., Mandt, K. E., Cooper, P. D., Umurhan, O. M., & Chevrier, V. F. (2018). Recent Advancements and Motivations of Simulated Pluto Experiments. *Space Science Reviews*, *214*(8), 130.

Bargery, A. S., Lane, S. J., Barrett, A., Wilson, L., & Gilbert, J. S. (2010). The initial responses of hot liquid water released under low atmospheric pressures: Experimental insights. *Icarus*, *210*(1), 488–506.

Birch, S. P. D., Hayes, A. G., Poggiali, V., Hofgartner, J. D., Lunine, J. I., Malaska, M. J., Wall, S., Lopes, R. M. C., & White, O. (2019). Raised rims around Titan's sharp-edged depressions. *Geophysical Research Letters*, *46*(11), 5846–5854.

Bol'shutkin, D. N., Gasan, V. M., & Prokhvatilov, A. I. (1971). Temperature dependence of the crystalline lattice parameter of methane in the range of 11–70°K. *Journal of Structural Chemistry*, *12*(4), 670–672.

Brož, P., Krýza, O., Patočka, V., Pěnkavová, V., Conway, S. J., Mazzini, A., Hauber, E., Sylvest, M. E., & Patel, M. R. (2023). Volumetric changes of mud on Mars: Evidence from laboratory simulations. *Journal of Geophysical Research. Planets*, *128*(12). https://doi.org/10.1029/2023je007950

Brož, Petr, Krýza, O., Wilson, L., Conway, S. J., Hauber, E., Mazzini, A., Raack, J., Balme, M. R., Sylvest, M. E., & Patel, M. R. (2020). Experimental evidence for lava-like mud flows under Martian surface conditions. *Nature Geoscience*, *13*(6), 403–407.

Buldovicz, S. N., Khilimonyuk, V. Z., Bychkov, A. Y., Ospennikov, E. N., Vorobyev, S. A., Gunar, A. Y., Gorshkov, E. I., Chuvilin, E. M., Cherbunina, M. Y., Kotov, P. I., Lubnina, N. V., Motenko, R. G., & Amanzhurov, R. M. (2018). Cryovolcanism on the Earth: Origin of a Spectacular Crater in the Yamal Peninsula (Russia). *Scientific Reports*, *8*(1), 13534.

Cruikshank, D. P., Roush, T. L., Owen, T. C., Geballe, T. R., de Bergh, C., Schmitt, B., Brown, R. H., & Bartholomew, M. J. (1993). Ices on the surface of triton. *Science*, *261*(5122), 742–745.

Davis, J. A., Rodewald, N., & Kurata, F. (1962). Solid-liquid-vapor phase behavior of the methane-carbon dioxide system. *AIChE Journal. American Institute of Chemical Engineers*, *8*(4), 537–539.

Engle, A. E., Hanley, J., Dustrud, S., Thompson, G., Lindberg, G. E., Grundy, W. M., & Tegler, S. C. (2021). Phase diagram for the methane–ethane system and its implications for titan's lakes. *The Planetary Science Journal*, *2*(3), 118.

Fagents, S. A. (2003). Considerations for effusive cryovolcanism on Europa: The post-Galileo perspective. *Journal of Geophysical Research*, *108*(E12). https://doi.org/10.1029/2003je002128
25

Howard, A. D., Moore, J. M., Umurhan, O. M., White, O. L., Singer, K. N., & Schenk, P. M. (2023). Are the surface textures of Pluto's Wright Mons and its surroundings exogenic? *Icarus*, *405*, 115719.

Howard, A. D., Moore, J. M., White, O. L., Umurhan, O. M., Schenk, P. M., Grundy, W. M., Schmitt, B., Philippe, S., McKinnon, W. B., Spencer, J. R., Beyer, R. A., Stern, S. A., Ennico, K., Olkin, C. B., Weaver, H. A., & Young, L. A. (2017). Pluto: Pits and mantles on uplands north and east of Sputnik Planitia. *Icarus*, *293*, 218–230.

Istomin, V. A., Chuvilin, E. M., Sergeeva, D. V., Bukhanov, B. A., Badetz, C., & Stanilovskaya, Y. V. (2020). Thermodynamics of freezing soil closed system saturated with gas and water. *Cold Regions Science and Technology*, *170*, 102901.

Jez-dotowski, A., Misiorek, H., Sumarokov, V. V., & Gorodilov, B. Y. (1997). Thermal conductivity of solid methane. *Physical Review. B, Condensed Matter*, *55*(9), 5578–5580.

Kohn, J. P., & Kurata, F. (1958). Heterogeneous phase equilibria of the methane—hydrogen sulfide system. *AIChE Journal. American Institute of Chemical Engineers*, *4*(2), 211–217.

Leibman, M., Kizyakov, A., Plekhanov, A. V., & Streletskaya, I. (2014). New permafrost feature – dep crater in central Yamal (west Siberia, rusia) as a response to local climate fluctuations. *Geography, Environment, Sustainability*, *7*, 68–79.

Litwin, K. L., Zygielbaum, B. R., Polito, P. J., Sklar, L. S., & Collins, G. C. (2012). Influence of temperature, composition, and grain size on the tensile failure of water ice: Implications for erosion on Titan. *Journal of Geophysical Research*, *117*(E08013). https://doi.org/10.1029/2012JE004101

Lorenz, R. D., McKay, C. P., & Lunine, J. I. (1997). Photochemically driven collapse of Titan's atmosphere. *Science*, *275*(5300), 642–644.

Lugo, R., Fournaison, L., & Guilpart, J. (2006). Ice–liquid–vapour equilibria of ammonia and ethanol aqueous solutions applied to the production of ice-slurries: Prediction and experimental results. *Chemical Engineering and Processing: Process Intensification*, *45*(1), 66–72.

Malaska, M. J., Schoenfeld, A., Wynne, J. J., Mitchell, K. L., White, O., Howard, A., Moore, J., & Umurhan, O. (2022). Potential caves: Inventory of subsurface access points on the surface of titan. *Journal of Geophysical Research. Planets*, *127*(11). https://doi.org/10.1029/2022je007512

Mancinelli, R. L., & Banin, A. (2003). Where is the nitrogen on Mars? *International Journal of Astrobiology*, *2*(3), 217–225.

Martin, C. R., & Binzel, R. P. (2021). Ammonia-water freezing as a mechanism for recent cryovolcanism on Pluto. *Icarus*, *356*, 113763.

Mastrogiuseppe, M., Hayes, A. G., Poggiali, V., Lunine, J. I., Lorenz, R. D., Seu, R., Le Gall, A., Notarnicola, C., Mitchell, K. L., Malaska, M., & Birch, S. P. D. (2018). Bathymetry and composition of Titan's Ontario Lacus derived from Monte Carlo-based waveform inversion of Cassini RADAR altimetry data. *Icarus*, *300*, 203–209.

Mastrogiuseppe, M., Poggiali, V., Hayes, A. G., Lunine, J. I., Seu, R., Mitri, G., & Lorenz, R. D. (2019). Deep and methane-rich lakes on Titan. *Nature Astronomy*, *3*(6), 535–542.
27